# Unsupervised Segmentation of B-Mode Echocardiograms

Melissa C. Brindise, Brett A. Meyers, Shelby Kutty, and Pavlos P. Vlachos

***Abstract*—** We present a method for unsupervised segmentation of echocardiograms (echo). The method uses an iterative Dijkstra's algorithm, a strategic node selection, and a novel cost matrix formulation based on intensity peak prominence and is thus termed the "Prominence Iterative Dijkstra's" algorithm, or ProID. Although the current analysis focuses on the left ventricle (LV), ProID is applicable to all four heart chambers. ProID was tested using artificial echo images representing five different systems. Results showed accurate LV contours and volume estimations as compared to the ground-truth for all systems. Subsequently, ProID was used to analyze a clinical cohort of 66 pediatric patients, including both normal and diseased hearts. Output segmentations, end-diastolic, end-systolic volumes, and ejection fraction (EF) were compared against manual segmentations from two expert readers. ProID maintained an average Dice similarity score of 0.93 when comparing against manual segmentation. Comparing the two expert readers, the manual segmentations maintained a score of 0.93 which increased to 0.95 when they used ProID. Thus, ProID successfully reduced the inter-operator variability across the two expert readers. Overall, this work demonstrates that ProID yields accurate boundaries across all age groups, disease states, and echo platforms with low computational cost, thereby establishing its clinical usefulness.

*Index Terms*—**Heart, Segmentation, Ultrasound.**

## I. INTRODUCTION

B-MODE echocardiography is the leading non-invasive imaging modality to examine left ventricle (LV) health due to its availability, affordability, and ease-of-use [1]. Physicians can assess LV shape and function using a B-mode apical four-chamber long-axis (A4C) cine recording, obtaining end-systolic volume (ESV), end-diastolic volume (EDV), stroke volume, ejection fraction (EF), and cardiac output measurements [1]. Such measurements require an accurate segmentation of the LV boundary. Currently, manual segmentation is the most commonly used clinical approach even though it introduces high inter- and intra-user variability, requires expert training, and is time-consuming to perform [2].

To mitigate these limitations and ensure repeatable measurements, semi- and fully-automated segmentation tools have been proposed for more than 30 years [3]–[5]. Contour detection algorithms fit deformable models or curves to intensity-based features, such as image edges. These are popular because of their ease of implementation and inexpensive computational cost. However, they have limited success as the transducer position, noise, and signal loss can corrupt segmented boundaries [3], [4]. Advancements towards active contour methods have been proposed to overcome these limitations. Skalski et al. [6] incorporated the Hough transform and Sarti et al. [7] employed probability models into the energy function, both seeking to avoid intensity-based features which are more sensitive to noise and signal attenuation.

Shape priors and tissue motion tracking have been used to improve contour detection algorithms. Shape priors, like those used by Dietenbeck et al. [8] and Zhou [9], use features that describe an expected shape. As algorithms deform the model, a set of penalties are applied based on the expectation. Dydenko et al. [10] used priors from tracking tissue motion and physician-provided delineations to create a multi-frame contour detection algorithm. In general, these tools are reliable, reporting similarities to trained clinicians of 92%-97% [8].

Deep learning segmentation methods have also been used. Unlike shape prior methods, these tools perform unsupervised learning based on image information. Carneiro et al. [11] used a two-stage, deep neural network which first selects regions from test images where the LV is fully present, then automatically segments the LV contour. However, the network was limited by the training set, which used only 400 healthy and 12 diseased LV patients. In 2015, U-Net convolutional neural networks (CNN) were first reported for use with medical image segmentation [12]. Studies using CNNs have since validated [13], further developed [14], [15], and demonstrated the method on large datasets [14], [16]. Large dataset results showed average similarity scores to expert delineations from 89% to 94%. Moreover, Zhang et al. [16] showed good results of the method across multiple echo systems. However, to achieve high accuracy, both shape priors and deep learning methods require large training sets with expert segmentations, which are generally difficult to obtain.

In general, high computational costs and the need for large training datasets are major limitations of existing automated segmentation methods, and hinder the clinical usefulness and

M. C. Brindise and B.A. Meyers are with the Department of Mechanical Engineering, Purdue University, 585 Purdue Mall, West Lafayette, IN, 47907, USA (e-mail: mbrindis@purdue.edu; meyers18@purdue.edu).

S. Kutty is with The Johns Hopkins University School of Medicine. He is the Director of The Helen B. Taussig Heart Center, M2315, 1800 Orleans St, Baltimore, MD, 21287, USA (e-mail: skutty1@jhmi.edu).

P. P. Vlachos is with the Department of Mechanical Engineering and the Weldon School of Biomedical Engineering, Purdue University, 585 Purdue Mall, West Lafayette, IN, 47907, USA (e-mail: pvlachos@purdue.edu).

adaptability of such methods. Herein, we introduce a new unsupervised LV segmentation framework which overcomes these gaps by using a modified Dijkstra's algorithm that does not require shape or temporal priors or any training. Our approach relies on a novel feature-based weighting of the shortest-path cost function to yield improved robustness and computation cost. We tested our method using artificial echocardiogram images [17] with predefined, "ground truth" boundaries. Clinical validation was done using a 66-patient cohort with 42 healthy individuals, 20 with hypertrophic cardiomyopathy (HCM), and 4 with dilated cardiomyopathy (DCM). Unsupervised measurements of ESV, EDV, and EF were compared against two manual expert readers.

## II. Algorithm

Unsupervised segmentation methods aim to find a boundary that connects the endpoints of the mitral annulus and optimally represents the LV in a cardiac echo scan. Based on the representation of the LV, the optimal boundary should have a smooth pathline and contain primarily high-intensity pixels. Dijkstra's algorithm is well-suited to identify this path in polar unwrapped scan images. Dijkstra's algorithm is a graph-based procedure that identifies the lowest "cost" path through a node network (i.e image pixels) spanning between a start and endpoint (i.e., the mitral annulus points). However, in its generic form, Dijkstra's algorithm is highly sensitive and often fails due to the signal dropout, LV clipping, and image noise echo scans often contain. Further large scan image sizes yield impractically high computational costs of Dijkstra's algorithm. Therefore, we propose a modified Dijkstra's algorithm which uses a strategic node selection, a novel cost formulation, and an iterative implementation for LV segmentation.

Our algorithm is initialized by three user-input point (UIP) selections on an arbitrary scan image (Figure 1a). The three UIPs are tracked in time using a standard cross-correlation with a 256 x 256-pixel window. For the LV, these three UIPs correspond to the apex and two ends of the mitral annulus. The LV center $(x_c, y_c)$ is defined as the geometric center of a triangle formed by the three UIPs and is used to transform the images from Cartesian to polar coordinates $(r, \theta)$ according to:

$$x_o = x - x_c,$$
$$y_o = y - y_c,$$
$$r = \sqrt{x_o^2 + y_o^2},$$
$$\theta = atan2(y_o, x_o) \quad (1)$$

where $x_o$ and $y_o$ are the LV-centered (x, y) and $atan2$ is:

$$atan2(y, x) = \arctan\left(\frac{y}{x}\right) + \frac{\pi}{2} sign(y) * (1 - sign(x)). \quad (2)$$

### A. Modified and Iterative Dijkstra's Algorithm

#### 1) Basics of Dijkstra's Algorithm

An illustrative example of Dijkstra's algorithm is given in supplemental Figure S1. For Dijkstra's algorithm, a node network including neighborhood connections and node cost (NC) is first defined. From the start node, the algorithm iteratively marches through the network, "visiting" one node per iteration until the end node is visited. Each node is initially assigned a "tentative distance" (*TD*) of infinity, except for the starting node, which is assigned a TD of zero. The node TD represents the cost of a path as it passes through that node. From the current node, $n_o$, the cost of the path to travel to each neighboring node, $n_n$, is evaluated as $TD(n_n)_{poss}$. If $TD(n_n)_{poss}$ is less than the neighboring node's current TD, the TD is reassigned as $TD(n_n)_{poss}$. The total path cost typically incorporates the summation of node costs (NC) which it passes through and a 'path distance cost' or pathlength. If path smoothness is of interest, a 'path angle cost', based on the angle between nodes, can be included. After each iteration, the next node visited is determined as the one having the lowest TD among the unvisited nodes.

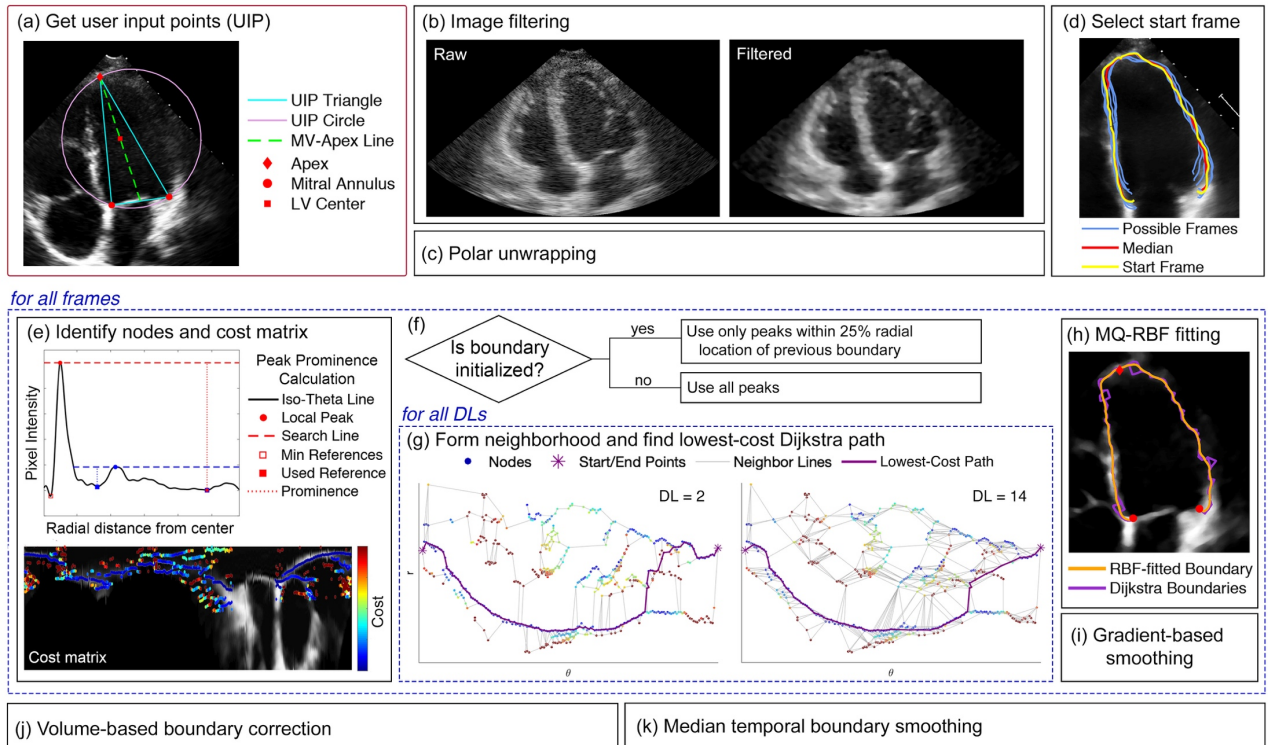

Figure 1. Flow-chart of the proposed LV segmentation method, ProID. Red boxes indicate user inputs. Blue dashed boxes indicate iterative loops. (a) User-input points (UIP) on raw scan image. (b) Raw and image filtered scan images. (c) Polar unwrapping. (d) Selection of start frame, t0. (e) Peak prominence calculation method with the first local peak calculation marked in red and the second local peak calculation marked in blue. Note the maximum-valued minimum reference is used in the prominence calculation. Cost matrix for example scan image according to Eq. 3. (f) Demonstration of nodes to use for Dijkstra's algorithm based on boundary initialization. (g) Iterative Dijkstra's algorithm neighborhood examples for a DL=2 and DL=14. (h) MQ-RBF boundary fitting of iterative Dijkstra's algorithm boundaries. (i) Gradient-based smoothing (see Figure 2). (j) Volume-based boundary correction (see Figure 3). (k) Median temporal boundary smoothing.

*2) Strategic node selection and cost formulation*

Using the unwrapped polar images, peaks of image intensity along the iso-theta lines (columns of the image in Figure 1e) are identified. Only these peak-based nodes within a radial distance of 1.5x the radius of the circle defined by the UIPs are included. This strategic node selection reduces the node network size by two orders of magnitude, from typically $10^5$ nodes to $10^3$ nodes.

For each node, the peak prominence and peak pixel intensity are used for the cost evaluation. Peak prominence quantifies the significance of the node peak intensity relative to other node peaks in the iso-theta line, providing an "uncertainty" for each peak. Peak prominence is computed as the difference between the peak height and a local minimum on the iso-theta line, as depicted in Figure 1e. The novel NC is then formulated as:

$$\text{NC} = (\hat{P} + \hat{I})^{-1} \quad (3),$$

where $\hat{P}$ and $\hat{I}$ are the normalized prominence and intensity values, respectively, and are normalized by their respective mean values across all nodes. Nodes with a cost greater than 1.25 times the mean NC of all nodes were discarded as they predominantly add noise and computational cost to the node neighborhood network. A path distance cost, $C_{dist}$, and path angle cost, $C_{angle}$, are defined according to:

$$C_{dist} = \sqrt{r_{n_n}^2 + r_{n_o}^2 - [2 * r_{n_n} * r_{n_o} * (\theta_{n_n} - \theta_{n_o})]} \quad (4),$$

$$C_{angle} = 0.1 * |90° - \text{atan2}(\theta_{n_n} - \theta_{n_o}, r_{n_n} - r_{n_o})| \quad (5),$$

where $atan2$ is the function defined in Eq. 2, $(r_{n_o}, \theta_{n_o})$ is the location of the currently visited node $n_o$, $(r_{n_n}, \theta_{n_n})$ is the location of the neighboring node $n_n$, and Eq. 4 is evaluated in pixels. Eq. 5 is evaluated in degrees and the '0.1' factor is used to bring NC, $C_{dist}$, and $C_{angle}$ to the same order of magnitude. $TD(n_n)_{poss}$ is evaluated according to:

$$TD(n_n)_{poss} = TD(n_o) + \alpha * NC_{n_n} + \gamma * C_{angle} + (C_{dist})^\beta \quad (6),$$

where $NC_{n_n}$ is the cost of the neighboring node (Eq. 3), and α, β, and γ are constants. The constant β enables exponential penalization of large path jumps. From testing, a β of 1.75 is effective for preventing segmentations from cutting through the LV. α and γ are normalizing factors which control the relative weight of the NC vs. path angle cost vs. path distance cost. Because the Dijkstra's algorithm is highly sensitive to the range and distribution of the cost matrix, proper setting of these constants is critical. For example, weighting the path angle cost too high biases paths towards a straight line in the polar domain (circle in Cartesian coordinates). For our purposes, a ratio of 10:5:1 of NC: $C_{dist}$: $C_{angle}$ proved optimal. This ratio was achieved using an α of 15 and a γ of 0.2. For other applications the optimal ratio of these three terms would need to be explored.

*3) Iterative Implementation*

Although strategic node selection reduces the dimension of the problem, it produces a non-structured grid of nodes which complicates the neighborhood definition. Moreover, with noise and signal dropout, it is at times desirable for paths to "jump" over iso-theta lines with no or noisy nodes. Thus, adjusting the allowable theta-distance limit (DL) between neighboring nodes can yield different paths, as illustrated in Figure 1g. However, an "optimal" DL for a given image cannot be determined *a priori* because the presence of noise and image artifacts is challenging to detect. Thus, several iterations of the Dijkstra's algorithm are performed, spanning a range of DLs. For the LV, we used a DL ranging from 2° to 14° in increments of 2°, resulting in seven iterations. A DL greater than 14° sometimes yielded large boundary jumps that cut off parts of the LV, as a lower number of nodes in the path often yields a lower path cost. Each node is required to have at least one "forward neighbor", sometimes requiring an exception to the DL.

To ensure the boundary passes through the LV apex, Dijkstra's algorithm is run on each boundary half (from the apex to each mitral annulus point) separately. From the seven unique boundaries, a single boundary is fitted using a weighted multi-quadratic radial basis function (MQ-RBF), as shown in Figure 1h. The MQ-RBF weights are set as the mean $\hat{P}$ for each path.

*4) Boundary Initialization*

To ensure temporal boundary smoothness, reduce the likelihood of image artifacts corrupting boundaries, and minimize computational cost, the nodes used in the modified Dijkstra's algorithm should be initialized by the previous time step's boundary when possible. For initialization, an expected boundary for the current time step is computed by displacing the previous boundary by the average radial displacement of the apex and MV points between the two time steps. Only nodes within a 25% radial distance tolerance of the expected boundary are included in the node network. If the average change between the initialized and previous time step's boundaries is higher than 10%, this typically indicates the algorithm clipped LV regions, and the current frame is rerun uninitialized.

Using this initialized approach, ProID's selection of the first frame to evaluate, or "start frame", is important since it prescribes the remaining boundaries. If an image with high noise or image artifacts that corrupts the resulting boundary is used as the start frame, the boundary identification for the entire scan time-series will likely fail. To minimize the occurrence of such an issue, five possible starting frames around end diastole of the first recorded beat are evaluated. End diastole was chosen because it typically has the clearest scans with the lowest papillary muscle image intensity. Uninitialized boundaries are evaluated using each of the five possible start frames as well as using the median image of all time-steps. The time frame whose boundary is closest to the median of these boundaries is selected as the starting frame, as shown in Figure 1d.

Overall, the proposed algorithm incorporates an enhanced node selection criterion, peak prominence-based node cost assessment, a unique iterative implementation of Dijkstra's algorithm, and a temporally based initialization procedure. For brevity, we term this method the 'prominence-based iterative Dijkstra's' method, or ProID, for the remainder of the paper.

### B. LV-specific image and boundary smoothing

*1) Adaptive contrast enhancement of scan images*

Adaptive contrast enhancement (ACE) was applied to the raw scan images to account for varying contrast-to-noise ratios (CNR) across echo systems and produce evaluation images with similar pixel intensity distributions. ACE scan images, $Im_{ACE}$, are computed according to:

$$Im_{ACE} = \frac{Im_{raw} - I_1}{I_2 - I_1} \quad (7)$$

where $Im_{raw}$ is the raw scan image, and $I_1$ and $I_2$ are one-half the mean and the maximum pixel values, respectively, along the MV-apex line, which connects the apex and the center of the MV (Figure 1a). Subsequently, an 11 x 11-pixel median filter is used to smooth the ACE-enhanced images. Figure 1b shows a raw and a filtered scan image.

*2) Gradient-based and temporal boundary smoothing*

Gradient-based smoothing (Figure 2) was applied to the MQ-RBF boundaries. First, a boundary angle between adjacent boundary points in Cartesian coordinates is evaluated. The smoothing seeks to remove peaks in the boundary angle array which typically indicate non-physical path oscillations. For each angle peak, a small surrounding segment of the boundary is extracted. Each segment is either bounded by adjacent boundary angle peaks or limited to consist of at most 8% of the surrounding boundary, whichever criterion produced a smaller segment size. For each segment, a set of alternative possible segments are formed by varying the location of a control point on the segment corresponding to the angle peak, as shown in Figure 2b. The path segment which optimally balances the normalized evaluation criteria of shortest pathlength, lowest number of angle peaks, lowest maximum derivative magnitude, and highest average image pixel intensity of points contained on the segment replaces the original extracted boundary section. Two iterations of this gradient-based smoothing were completed for each boundary. Figure 2c shows an example boundary after gradient smoothing.

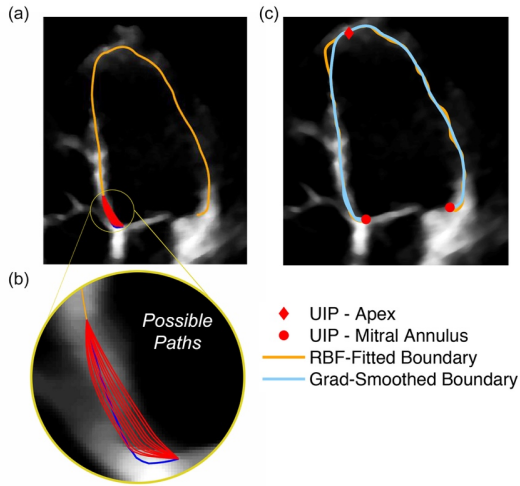

Figure 2. Gradient-based smoothing. (a) Boundary and small segment of code being evaluated. (b) Set of possible paths evaluated. (c) Final gradient-smoothed boundary.

*3) Volume-based boundary correction*

To account for varying clinician segmentation preferences, a volume-based boundary correction step is applied to the gradient-smoothed boundaries. For each frame, inner and outer boundaries are first identified using a line perpendicular to the smoothed boundary, as shown in Figure 3a. The outer boundary is defined as the first point radially outwards, where the pixel intensity drops below 67% of that of the boundary point. The inner boundary is defined as the most radially inwards point of either the location of a first minimum intensity peak inward of the boundary or the first point whose pixel intensity is less than 25% of the smoothed boundary pixel intensity. A lower threshold was used for the inner boundary so that it properly captured the papillary muscles. The inner and outer boundary points are filtered using the discrete wavelet transform applied in polar coordinates and then smoothed using the gradient-based and median temporal smoothing. Figure 3b and 3c show example inner and outer points and boundaries.

The position of each point along the original boundary, $r_b$, relative to the inner ($r_i$) and outer ($r_o$) boundaries, or a "relative boundary position (RBP)", is then computed according to:

$$RBP = \frac{r_b - r_i}{r_o - r_i} \quad (8),$$

and can be interpreted as:

$$RBP = \begin{cases} > 1 & r_b \text{ is radially outwards of } r_o \\ 1 & r_b \text{ is located at } r_o \\ 0 & r_b \text{ is located at } r_i \\ < 0 & r_b \text{ is radially inwards of } r_i \\ otherwise & r_b \text{ is radially between } r_o \text{ and } r_i \end{cases} \quad (9).$$

For each frame, the mean RBP (MRBP) is computed and is considered the "raw MRBP". A desired corrected MRBP (cMRBP) is then assigned to each scan frame. The cMRBP is initially set, as shown in Figure 3d, by interpolating the raw LV volume array (Figure 3f) where the average EDV and ESV correspond to cMRBP values of 0.47 and 0.27, respectively. The systolic cMRBP is set lower because papillary muscles maintain higher pixel intensity at systole, which lower the peak prominence computed for the region. This often pushes ProID boundaries radially outward, while clinical boundaries are typically drawn radially inside the papillary muscles. It is important to note that these values were arbitrarily set as well-suited values for the ProID testing herein. Future work should aim to systematically determine optimal values. The initial cMRBP is then adjusted to account for disparities between each beat's diastolic and systolic volumes, as observed in Figure 3f. The beat variability adjusted cMRBP is denoted as 'cMRBP-BV' in Figure 3d. The cMRBP (denoted as 'cMRBP-In/Out' in Figure 3d) is then adjusted to account for differences in the area between the inner and outer boundaries across frames (Figure 3e) which can produce a bias in the scaling. 'cMRBP-In/Out' is used for the boundary scaling.

The volume-based correction is done by scaling the original boundary such that the MRBP of the corrected boundary matches its assigned cMRBP. For example, Frame 1 of the scan shown in Figure 3d the MRBP should be adjusted from 0.69 to 0.53 by moving the boundary radially inward. A power-six scaling is used to adjust the RBP of each boundary point, so points furthest from the cMRBP are corrected most. Figures 3c and 3f show the volume-corrected boundary and curve.

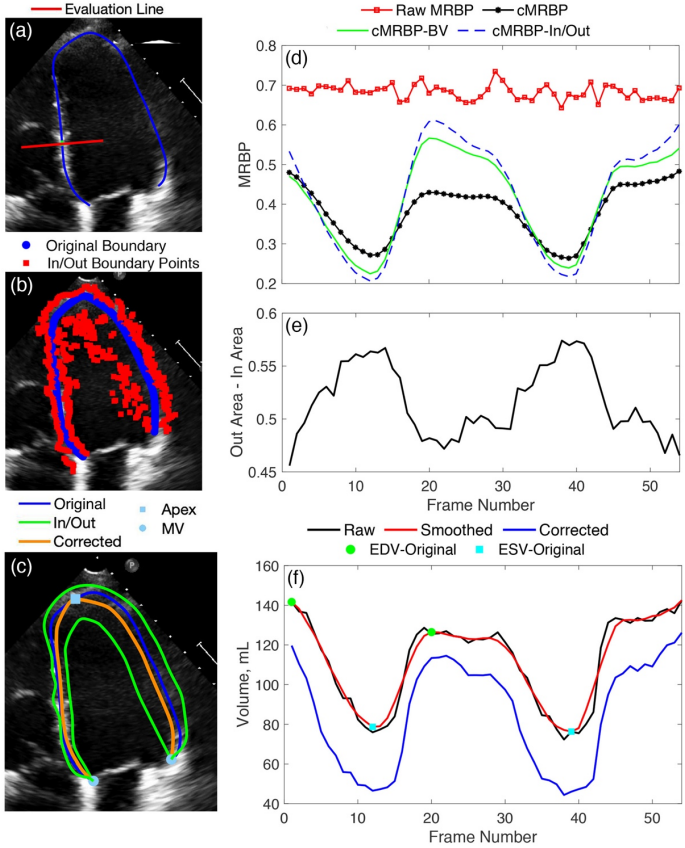

Figure 3. Schematic of the volume-based boundary correction method. (a) Lines are drawn perpendicularly to the boundary to (b) identify the inner and outer boundary points. (c) Original smoothed boundary, inner and outer boundaries, and corrected boundary for a given scan. (d) Mean relative boundary position (MRBP). (e) Difference in area between inner and outer boundaries across scans. (f) Raw, smoothed, and corrected LV volumes.

## III. ANALYSIS AND TESTING

### A. Synthetic Ultrasound

ProID was tested using synthetic LV A4C echos [17], [18] generated for the European Association of Cardiovascular Imaging - American Society of Echocardiography (ASE) Task Force [19]–[23]. The synthetic echos modeled GE Vivid E9, Hitachi Prosound $\alpha 7$, Philips iE 33 Vision, Siemens SC2000, and Toshiba Artida ultrasound systems. Each dataset provided corresponding ground truth boundaries.

### B. Clinical Echocardiogram Study

We evaluated the clinical capabilities of ProID using a retrospective study of pediatric heart failure patients and with age-matched healthy controls collected at the University of Nebraska Medical Center (Omaha, Nebraska, USA). The cohort consisted of 66 subjects: 4 dilated cardiomyopathy (DCM), 20 hypertrophic cardiomyopathy (HCM), and 42 age-matched controls (Normal). Cohort demographics are provided in Table 1 and function indices are provided in Table 2.

Each subject underwent a routine echocardiogram with an iE33 ultrasound system (Philips Healthcare, Andover, MA, USA), per the ASE guidelines [24]. Doppler echocardiography measurements were collected in the A4C view. Cine recordings were stored in Digital Imaging and Communications in Medicine (DICOM) format.

Two experienced sonographers manually segmented EDV and ESV for all available recorded beats from all subjects (373 total beats). ProID was run using UIPs extracted from the first-beat manual segmentations of each reader as well as using points selected by an untrained reader.

Table 1. Clinical cohort demographics

| Characteristics | Normal ($n = 42$) | DCM ($n = 4$) | HCM ($n = 20$) |
|---|---|---|---|
| Age (years) | 17.98 ± 8.86 | 14.50 ± 6.24 | 18.74 ± 10.47 |
| BSA (m$^2$) | 1.66 ± 0.56 | 1.52 ± 0.64 | 1.81 ± 0.69 |
| Height (cm) | 159.25 ± 29.11 | 147.90 ± 57.60 | 159.81 ± 30.86 |
| Weight (kg) | 63.50 ± 30.42 | 57.60 ± 35.65 | 75.71 ± 41.03 |
| Heart Rate (bpm) | 67.47 ± 17.26 | 92.50 ± 33.81 | 72.88 ± 18.78 |

Table 2. Indices for LV dimensions and functional parameters.

| *Ventricular Dimensions* | Normal ($n = 42$) | DCM ($n = 4$) | HCM ($n = 20$) |
|---|---|---|---|
| End Diastolic Volume (ml) | 98.85 ± 39.19 | 178.75 ± 83.92 | 96.28 ± 37.66 |
| End Systolic Volume (ml) | 37.80 ± 15.84 | 117.75 ± 60.31 | 36.03 ± 18.40 |
| Stroke Volume (ml) | 61.29 ± 24.39 | 61.00 ± 33.32 | 59.44 ± 21.24 |
| Ejection Fraction (%) | 62.16 ± 3.50 | 34.25 ± 14.93 | 63.06 ± 6.01 |
| *Functional Parameters* | | | |
| E-wave velocity (cm/s) | 82.20 ± 19.70 | 102.25 ± 29.80 | 83.19 ± 19.59 |
| A-wave velocity (cm/s) | 42.84 ± 10.95 | 61.25 ± 34.74 | 61.13 ± 33.65 |
| e' velocity (cm/s) | 17.58 ± 3.22 | 11.38 ± 2.63 | 10.51 ± 3.04 |
| E/A ratio | 2.00 ± 0.59 | 2.24 ± 1.62 | 1.55 ± 0.50 |
| E/e' ratio | 4.06 ± 1.21 | 8.22 ± 4.30 | 7.34 ± 2.58 |

### C. Quantitative Evaluation

The Dice similarity score (DSS) was used to evaluate segmentation overlap and thus accuracy. DSS is defined as:

$$DSS = \frac{2(X_{test} \cap X_{reference})}{X_{test} + X_{reference}}, \quad (10)$$

where $X_{test}$ is the algorithm output segmentation and $X_{reference}$ is a trusted segmentation. Dice similarity score ranges from 0, no overlap exists, to 1, the segmentations fully overlap.

Synthetic echocardiogram segmentation accuracy was assessed using the Dice similarity score. Values are reported from a 1000-trial Monte-Carlo simulation that varied the input positions based on the distance from the inner boundary to the outer boundary of the ground truth. The mean and standard deviation for each quantity were calculated as a function of the image contrast-to-noise ratio (CNR), defined as

$$CNR = \frac{|\mu_{cavity} - \mu_{myocardium}|}{\sqrt{\sigma^2_{cavity} + \sigma^2_{myocardium}}}, \quad (11)$$

where $\mu_{cavity}$ and $\mu_{myocardium}$ are the mean pixel intensity inside the LV cavity and myocardium, and $\sigma^2_{cavity}$ and $\sigma^2_{myocardium}$ are the corresponding variance of pixel intensity.

Clinical echocardiogram accuracy was assessed using EDV, ESV, and EF, which is computed according to:

$$EF\ (\%) = \frac{EDV - ESV}{EDV}. \quad (12)$$

## IV. RESULTS

### A. Synthetic Echocardiogram Analysis

Figure 4 provides the segmented boundaries identified for the simulated data from each of the five systems. Each system maintains a varying CNR, which often affects unsupervised segmentation algorithms. Both the segmentation and volume curves show ProID was accurate across all systems, indicating the robustness of the ACE step. All volume curves were within the ground-truth range (between the epi- and endocardium). Further, all segmented boundaries were contained within the ground-truth region, except for the Hitachi system. Hitachi, which maintained a mid-range CNR of 3.3 (i.e. myocardium signal is 3.3x stronger than cavity signal), had some signal dropout along the lateral wall causing the boundary to be overestimated in this region. GE and Philips systems maintained the lowest CNR values of, on average, 1.8 and 2.3, respectively. For both of these systems, the volumes were within the mid-range ground truth region. Siemens and Toshiba maintained the highest CNR of 3.8 and 5.2, respectively. For these two cases, the measured volumes were closer to the endocardium. These results demonstrate that ProID boundaries trend towards the endocardium as the CNR increases.

Figure 5 shows the Dice Similarity score as a function of CNR. Dice similarity scores range from 0.82 ± 0.02 at a CNR of 1 to 0.96 ± 0.01 at a CNR of 7. Similarity scores improve with increasing signal strength, as expected. Nonetheless, ProID still yielded accurate boundaries at all CNR values.

### B. Clinical Echocardiogram Analysis

Figure 6 illustrates ProID-segmented boundaries compared to Reader-segmented boundaries for three patients across all three diagnoses. For each scan, three ProID boundaries are shown: one using the UIPs from Reader #1, one using the UIPs from Reader #2, and one using the UIPs from the paper author which we refer to as an "untrained reader". In all scans, only minor differences are observed between the ProID boundaries and volume trends, highlighting ProID's robustness to small variations in UIPs.

For the normal patient, both expert readers had EDV measurements of about 147 mL, with largely similar end-diastole contours (Figure 6a-1). The ProID diastolic boundaries displayed good qualitative agreement with both readers. ProID measured an EDV between 158 mL and 170 mL for all reader UIPs, a 7.5% to 15.7% difference from the manual drawings. At end-systole (Figure 6a-2) larger differences across expert readers are observed with Readers #1 and #2 measuring 98 mL and 67 mL, respectively. ProID measured volumes between 76 mL and 80 mL for all UIPs, an 11.9% to 18.4% difference from the manual drawings. In the normal patient systolic segmentations from Reader #2 and ProID, a notable difference in the apex location is observed. This can occur because the ultrasound is gated to begin scanning at diastole so the apex UIP is entered for diastole and tracked through time, while the reader re-selects the apex location between systole and diastole.

The HCM patient (Figure 6b) presents with septal thickening near the apex. As a result, the LV takes on an atypical shape. Expert reader EDV measurements between manual (68 mL vs. 89 mL) and ProID (73 mL vs. 79 mL) as well as the untrained reader (84 mL) were in good agreement with less than 11% difference. Again, end-systole manual measurements (Figure 5b-2) showed greater differences across Readers in terms of both contours and ESV (Reader #1: 18 mL vs. Reader #2: 40 mL). ProID results more closely followed Reader #2, though some shape differences are still observed near the septal wall. ProID ESV measurements were 22 mL, 32 mL, and 31 mL for Reader #1, Reader #2, and the untrained reader, respectively.

For the DCM patient shown in Figure 6c, the ventricle distended. End-diastole (Figure 6c-1) manual Reader contours were similar with slight differences in the mitral valve position.

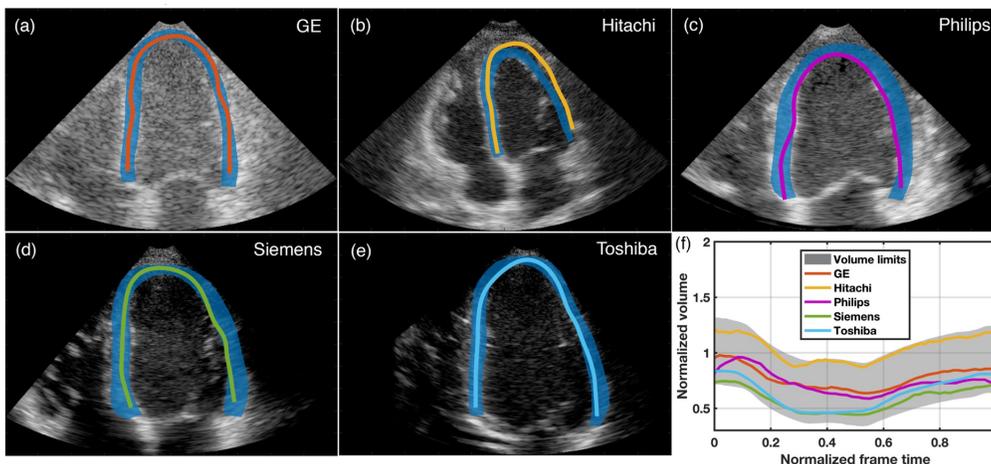
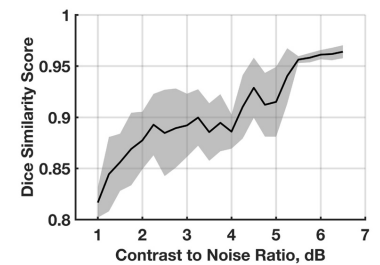

Figure 4. ProID segmented contours compared to "ground-truth" for artificial scans of (a) GE, (b) Hitachi, (c) Philips, (d) Siemens, and (e) Toshiba ultrasound systems. (f) Normalized boundaries computed from ProID for all ultrasound systems compared to ground-truth.

Figure 5. Dice similarity score as a function of contrast to noise ratio (CNR) across all artificial images from all vendors.

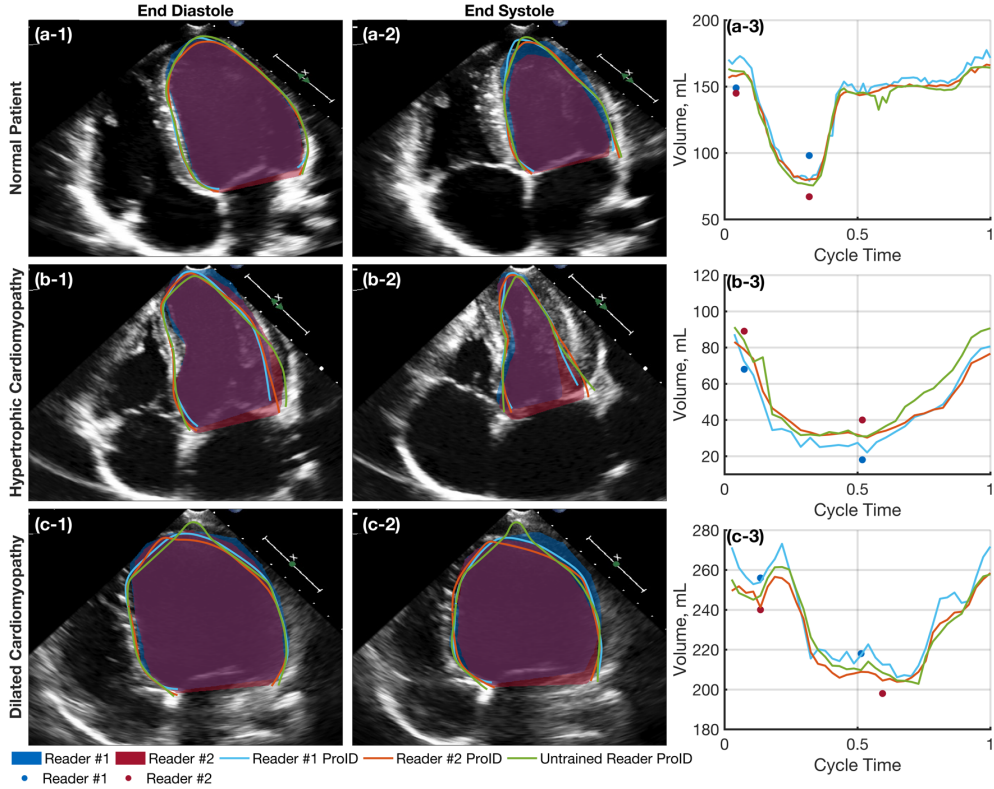

Figure 6. Demonstration of ProID on clinical images from a (a) normal subject, (b) HCM subject, (c) DCM subject at (-1) end diastole and (-2) end systole. (-3) Time-series LV volume curves from ProID as well as manual Reader measurements at end diastole and systole only.

Manual EDV measurements were 256 mL for Reader #1 and 240 mL for Reader #2. ProID contours slightly shortened near the apex and along the lateral wall where signal loss occurred, but volumes agreed with the expert manual measurements at about 254 mL for Reader #1, 242 mL for Reader #2, and 247 mL for the untrained reader. Similar trends are observed at end-systole (Figure 6c-2) with Reader #1 manually measuring 218 mL, while ProID measured 217 mL. Reader #2 manually measured 198 mL, while ProID using Reader #2 and the untrained reader UIPs both measured 205 mL.

Overall, Figure 6 demonstrates good agreement between expert manual contours and ProID contours across all three diagnoses. Table 3 provides the average Dice similarity scores comparing between the five segmentations (the two manual expert readers and ProID using UIPs from each reader as well as an untrained reader). Across all five comparisons, the average Dice Similarity scores were between 0.92-0.93 for the first recorded beat and 0.91-0.92 for the second recorded beat in each scan. The similarity between the first and second recorded beats demonstrates that ProID maintained accuracy even across longer scans. The average Dice Similarity scores for all Manual Reader comparisons were between 0.92-0.93 for normal patients, 0.88-0.90 for HCM patients, and 0.93-0.94 for DCM patients. HCM patients maintained slightly lower similarity scores largely due to the more abnormal LV shapes and smaller volumes. Comparing ProID using the three UIP implementations, Dice similarity scores of 0.95-0.96 were observed for Normal and DCM patients and 0.93-0.94 for HCM patients. Overall, the average scores were statistically similar, demonstrating ProID was robust across all readers.

Table 3. Dice similarity scores comparing all five segmentation methods (Manual expert Reader #1 and #2 as well as ProID with UIPs from Reader #1 and #2 and an untrained Reader). Scores were averaged based on recorded beat number and diagnosis. One-sigma standard deviations are also provided.

| Comparison | All Diagnoses | | All Beats | | |
|---|---|---|---|---|---|
| | Beat 1 | Beat 2 | Normal | HCM | DCM |
| ProID (Reader #1) vs. Manual Reader #1 | 0.93 ± 0.03 | 0.92 ± 0.04 | 0.93 ± 0.03 | 0.90 ± 0.04 | 0.93 ± 0.02 |
| ProID (Reader #2) vs. Manual Reader #2 | 0.93 ± 0.03 | 0.92 ± 0.05 | 0.93 ± 0.03 | 0.89 ± 0.05 | 0.95 ± 0.01 |
| ProID (Untrained) vs. Manual Reader #1 | 0.92 ± 0.03 | 0.91 ± 0.04 | 0.92 ± 0.03 | 0.88 ± 0.04 | 0.92 ± 0.02 |
| ProID (Untrained) vs. Manual Reader #2 | 0.92 ± 0.04 | 0.91 ± 0.05 | 0.92 ± 0.03 | 0.88 ± 0.05 | 0.94 ± 0.01 |
| Manual Reader #1 vs. Manual Reader #2 | 0.92 ± 0.03 | 0.91 ± 0.03 | 0.93 ± 0.03 | 0.89 ± 0.04 | 0.94 ± 0.02 |
| ProID (Reader #1) vs. ProID (Reader #2) | 0.95 ± 0.03 | 0.94 ± 0.03 | 0.95 ± 0.03 | 0.93 ± 0.03 | 0.95 ± 0.02 |
| ProID (Reader #1) vs. ProID (Untrained) | 0.95 ± 0.02 | 0.95 ± 0.03 | 0.95 ± 0.03 | 0.94 ± 0.03 | 0.95 ± 0.02 |
| ProID (Reader #2) vs. ProID (Untrained) | 0.96 ± 0.02 | 0.96 ± 0.02 | 0.96 ± 0.02 | 0.94 ± 0.03 | 0.96 ± 0.01 |

Figure 7 compares the EDV and ESV between both expert readers and ProID using all three sets of UIPs. In all cases, the performance was similar between the first (Figure 7-1) and second (Figure 7-2) recorded beats. Figure 7a compares ProID using UIPs from Reader #1 and Manual Reader #1 volumes, while Figure 7b compares volumes from ProID using UIPs from an untrained reader with those from Manual Reader #1. Figures 7c and 7d compare volumes from Manual Reader #2 with those from ProID using UIPs from Reader #2 and an untrained reader, respectively. About an 8.3 mL bias error was observed where ProID using Reader #1 UIPs underestimated volumes compared to Manual Reader #1. This error reduced to about 3.5 mL when using UIPs from an untrained reader. The Bland-Altman point distribution demonstrates no significant proportional error between ProID and Reader #1. Comparing ProID and Reader #2, ProID on average overestimated the volumes by 10.2 mL. A small proportional error is indicated where ProID further overestimated the volumes as the volume increased. For both Readers, similar ranges of the 95% limits of agreement (LoA) were observed for both UIP sets. ProID maintained a slightly lower agreement with Reader #2 than Reader #1, with the LoA increasing by about 7% for Reader #2.

Table 4 summarizes the average EDV and ESV for each of the five evaluation methods separated by diagnosis. In general, all ProID measurements for Normal and HCM patients are between those of Reader #1 and Reader #2. On average, ProID underestimated EDV compared to Reader #1 by 3.5% and 0.7% for Normal and HCM patients, respectively. Meanwhile, ProID overestimated EDV compared to Reader #2 by 6.1% for Normal patients and 14.9% for HCM patients. These differences almost doubled for ESV (Reader #1: 5.1% Normal, 2.1% HCM; Reader #2: 16.4% Normal, 35.7% HCM), while volumes reduced by 50% on average, confirming that the bias error discussed above was the dominant error source as opposed to a proportional error. For DCM patients, ProID slightly underestimated compared to both readers by on average 8.5% for diastole and 12.5% for systole.

Table 4. Average end-diastolic volumes (EDV) and end-systolic volumes (ESV) for each segmentation method across all three diagnoses.

|  |  | Manual | | ProID | | |
|---|---|---|---|---|---|---|
|  |  | Reader #1 | Reader #2 | Reader #1 | Reader #2 | Untrained |
| Normal | EDV | 132 ± 46 | 120 ± 40 | 126 ± 49 | 127 ± 49 | 129 ± 52 |
|  | ESV | 65 ± 25 | 53 ± 19 | 62 ± 27 | 61 ± 26 | 62 ± 26 |
| HCM | EDV | 153 ± 71 | 134 ± 52 | 148 ± 64 | 158 ± 70 | 156 ± 67 |
|  | ESV | 79 ± 54 | 57 ± 32 | 70 ± 43 | 81 ± 49 | 81 ± 48 |
| DCM | EDV | 205 ± 110 | 202 ± 111 | 177 ± 93 | 188 ± 96 | 193 ± 106 |
|  | ESV | 169 ± 100 | 151 ± 96 | 132 ± 76 | 139 ± 78 | 149 ± 93 |

Figure 8 compares the ejection fraction computed using ProID to those measured by Readers #1 and #2. Each box provides the statistical values for median, interquartile, and minimum and maximum ranges for EF measurements from the first (Figure 8-1) and second (Figure 8-2) beats of each scan. Figure 8a provides EF statistics for Normal subjects. First beat ProID EF measurements for Reader #1, Reader #2, and the untrained reader UIPs were 51.7%, 53.5%, 53.8%, respectively. Reader #1 and Reader #2 manual measurements (51.5% and 56.0%). The difference between ProID and manual measurements did not exceed 8%. Similarly, for the second beat, ProID median EF for Reader #1, Reader #2, and the untrained reader UIPs were 52.3%, 52.6%, 53.1%, respectively. Reader #1 and Reader #2 manual measurements were 52.0% and 55.3%, respectively. For the second beat, the max difference between ProID and manual Reader EF was 5%.

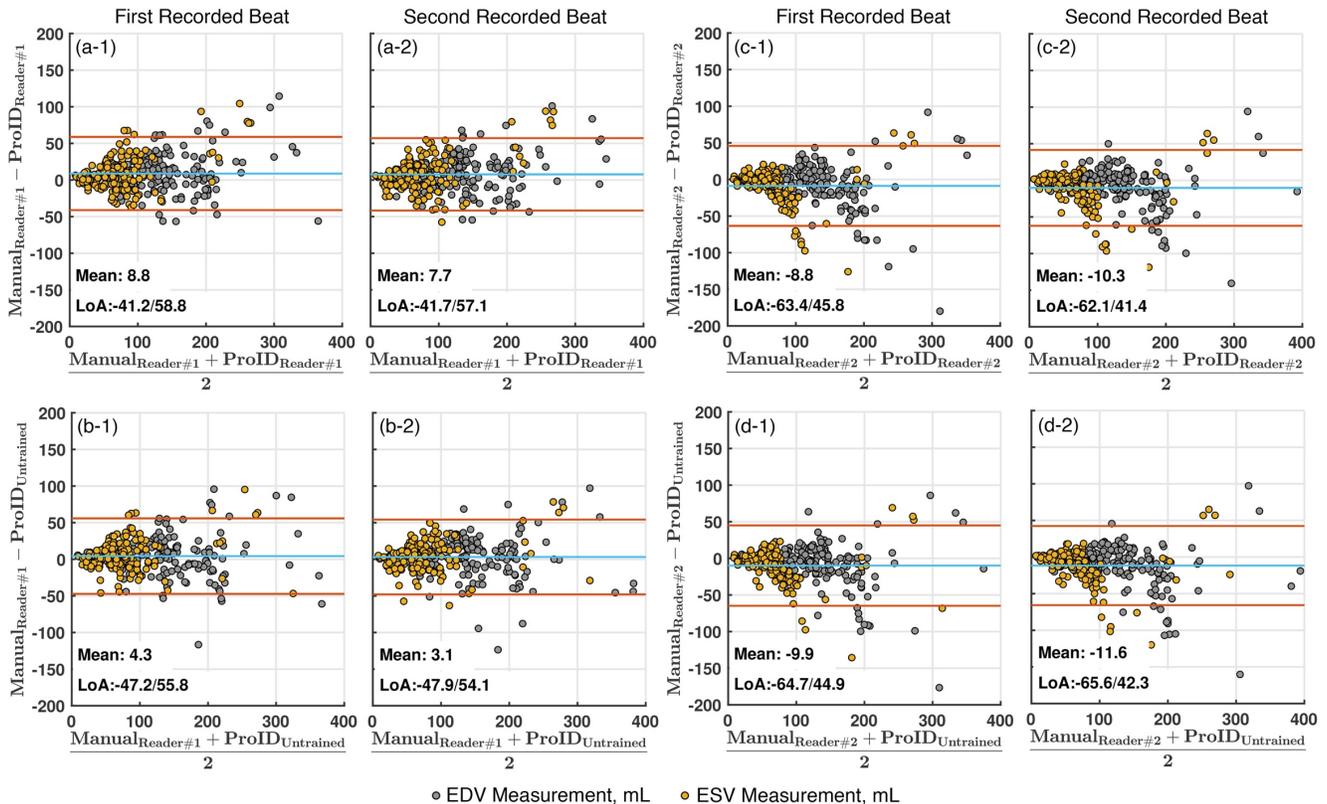

Figure 7. Bland-Altman analysis of end diastole volumes (EDV) and end systolic volumes (ESV). Comparisons of volumes between Manual Reader #1 and (a) ProID using UIPs from Reader #1 and (b) ProID using UIPs from an untrained reader. Comparisons of volumes between Manual Reader #2 and (c) ProID using UIPs from Reader #2 and (d) ProID using UIPs from an untrained reader. Analysis was separated by the first recorded beat (-1) and second recorded beat (-2) in each scan.

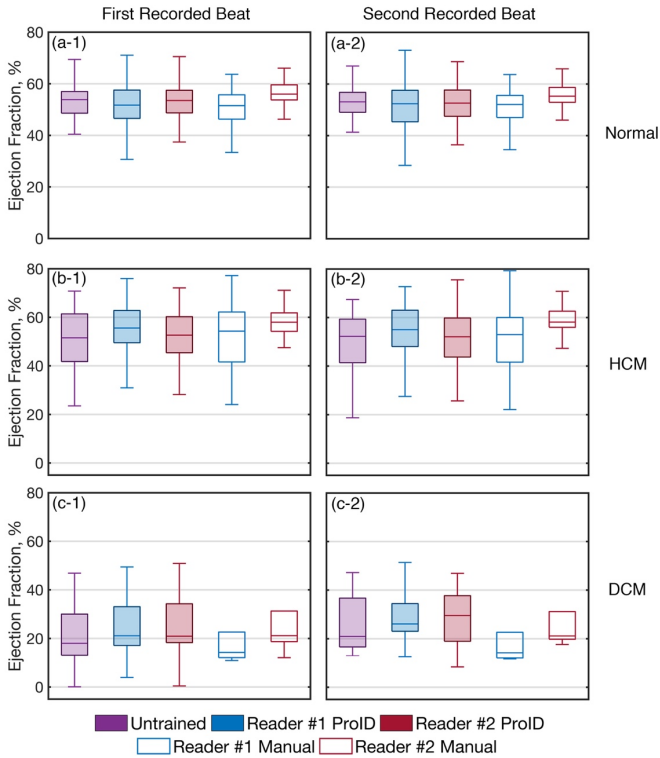

Figure 8. Box plots of ejection fraction (EF) for each of the five segmentation methods and each diagnosis. Median, interquartiles, minimum, and maximum EF are provided.

Figure 8b provides EF statistics of HCM subjects. ProID measurements for the first beat for Reader #1, Reader #2, and untrained reader UIPs were 55.6%, 52.6%, 51.5%, respectively. Manual measurements for Reader #1 and Reader #2 were 54.3% and 58.0%, respectively, for a max percent difference against ProID measurements of 11%. Second beat ProID EF measurements for Reader #1, Reader #2, and an untrained reader UIPs were 55.0%, 52.0%, 52.2%, respectively. Meanwhile, Reader #1 manually measured 53.0% and Reader #2 manually measured 58.1%. The max percent difference between ProID and manual measurements was 10%.

Finally, EF statistics of DCM patients are shown in Figure 8c. Larger discrepancies between beat 1 and beat 2 were observed due to smaller patient sample size (n=4). First beat ProID EF measurements for Reader #1, Reader #2, and an untrained reader UIPs were 21.1%, 21.0%, 18.0%, respectively. Reader #2 had a similar median EF measurement of 21.1%, while Reader #1 had a slightly lower median EF of 14.2%. However, the maximum percent difference between manual and ProID EF reached 49%. For the second beat, ProID EF medians for Reader #1, Reader #2, and the untrained reader UIPs were 26.0%, 29.5%, 20.9%, respectively, while manual median EF were unchanged. The maximum percent difference of beat 2 ProID compared to manual EF measurement was 108%.

## V. DISCUSSION

In this work, we introduced a novel unsupervised and automated segmentation algorithm, ProID, which was tested using both artificial and clinical LV B-mode echocardiograms. We demonstrated that ProID is robust across varying LV sizes and shapes and can accurately segment multi-beat recordings with just three user-input points. While Dijkstra's algorithm has been used previously to segment LVs ([25]), our iterative Dijkstra's algorithm implementation and novel peak prominence-based cost matrix employed in ProID provided robustness and computational efficiency.

Using 600x800 pixels B-mode scans and a standard CPU, ProID segments a frame in about 5-10 seconds, corresponding to typically less than 2 minutes per scan when parallel processed across 12 cores (using MATLAB). The peak prominence-based cost matrix method markedly improves computation time by reducing the number of nodes in the neighborhood which controls the computation time. Typically, methods using Dijkstra's algorithm use each image pixel as a node in the cost matrix. For the B-mode images used here, this corresponds to 480,000 nodes using the full image or about 90,000 nodes using a cropped region of interest. This is reduced to 2500 nodes by utilizing only the radial intensity peaks.

A major difficulty when developing any unsupervised segmentation method is the lack of a consensus among contours manually segmented by expert readers. Both inter- and intra-user variabilities are well-established limitations of manual LV segmentation. Segmentation contours from the two expert Readers used here, despite both readers working in the same hospital, differed by about 8% on average (according to Dice similarity scores), but could be much larger for certain frames. For example, in Figure 6a-2, there is a significant disparity between the apex location identified by Reader #2 and the apex location identified by Reader #1 (as well as ProID). For this frame, Reader #2 yielded an ESV 32% lower than that of Reader #1. On average, volume measurements across all diagnoses from Reader #2 were 8% lower at diastole and 19% lower at systole than that of Reader #1. Conversely, when using ProID, the average difference between volume measurements for the two expert readers dropped to 5% and 8% for EDV and ESV, respectively, a two-fold improvement. Even with UIPs entered from an untrained reader, ProID maintained, on average, a nearly zero difference in EDV and ESV for Normal and HCM patients and an 8% difference for DCM patients, as compared to using UIPs from both expert readers. Overall, this analysis highlights that ProID is able to reduce inter-user variability, yielding more standardized results across users.

A notable feature of ProID is that it does not presently use any machine learning (ML) algorithms while still maintaining similar Dice similarity scores as ML based algorithms (ProID: 0.92 vs. ML: 0.96 [14]). As a result, ProID requires fewer assumptions and is applicable across a broader scope of segmentation problems than such algorithms, which vastly increases its utility. Specifically, ML-based algorithms often excel for adult LV segmentation but struggle with fetal, neonatal, or pediatric LV segmentations, which most often have different sizes, shapes, and heart rates compared to adults. Conversely, ProID excels in all age groups, as demonstrated here where the clinical cohort contained both adult and pediatric scans. Moreover, because ProID does not rely on any training, it can be used to segment the right ventricle as well as both atria. ProID could also be adapted to segment boundaries outside of the heart and using other imaging modalities such as computed tomography or magnetic resonance imaging, notions to be explored in future work.

Some notable limitations of both ProID and this study exist. ProID relies on sufficient tissue signal to reliably segment the LV, which may not be available in very low-quality scans. Current machine learning methods perform more reliably in these cases. The iterative Dijkstra and MQ-RBF interpolant can overcome short signal dropouts, but this interpolation can lead to volume overestimation, and can ultimately fail for larger regions of signal dropout. Images with significant signal oversaturation can produce additional peaks, which may cause less accurate, low-cost paths to be found. The boundary volume correction used here is currently initialized using heuristic values and should be further studied and validated with additional data. For the simulated data used here, "ground-truth" segmentations were not based on expert clinical readers. Thus, conclusions on the analysis of the simulated data should be limited to ProID's performance relative to itself across the five ultrasound systems. In the clinical cohort, few dilated patients were available in the cohort, limiting analysis for this cardiomyopathy condition. Further, only two expert readers were used, precluding inter-user variability from being statistically analyzed and compared.

## ACKNOWLEDGMENTS

The authors thank Mary Craft for clinical scan collection. The authors also thank Benjamin Barnes, and Rita Long for their expert manual segmentations.